\documentclass{article}
\usepackage{amssymb,amsfonts,amsmath}
\usepackage{cite,enumerate,float}
\usepackage{color}
\usepackage{tikz}
\usetikzlibrary{arrows,snakes,backgrounds}
\DeclareGraphicsExtensions{.pdf,.png,.jpg}
\usepackage{multirow}
\usepackage{pgfplots}
\pgfplotsset{compat=1.9}
\usepackage{amsmath, amsfonts, amssymb, amsthm, mathtools}
\usepackage{rotating}
\usepackage{ytableau}
\ytableausetup{mathmode, boxsize=0.4 em}
\usepackage{braket}
\usepackage{float}
\usepackage{amsmath}
\usepackage{tikz}
\usetikzlibrary{calc,matrix}
\usepackage{wrapfig}
\usepackage{graphicx}
\graphicspath{{pictures/}}
\usepackage{hyperref}
\DeclareGraphicsExtensions{.pdf,.png,.jpg}
\usepackage{array}
\usepackage[T2A]{fontenc}
\usepackage[utf8]{inputenc}
\def\be{\begin{eqnarray}}
\def\ee{\end{eqnarray}}

\def\tr{{\rm tr}\,}

\usepackage{tikz}

\usepackage{xcolor}
\hypersetup{
    colorlinks=true,
    linkcolor=magenta,
    citecolor=blue,
    urlcolor=magenta
}

\definecolor{red}{rgb}{1,0,0}
\definecolor{orange}{rgb}{1,0.5,0}
\definecolor{violet}{rgb}{0.7,0,1}

\hoffset= - 1.0in

\begin{document}


\hfill MIPT/TH-20/26

\hfill IITP/TH-22/26

\hfill ITEP/TH-24/26

\bigskip

\centerline{\Large{\fontfamily{cmss}  \textbf{Character expansion of the Itzykson-Zuber pair correlator}}}

\bigskip
\bigskip

\renewcommand{\thefootnote}{\fnsymbol{footnote}}
\centerline{\textbf{Alexei Morozov$^{1,3,4,5}$\footnote[3]{morozov@itep.ru} \ and\ \   Hasib Sifat$^{1,2}$\footnote[9]{hasib.sifat@math.msu.ru}}}

\bigskip

\centerline{\it $^1$Moscow Institute of Physics and Technology, 141700, Dolgoprudny, Russia}
\centerline{\it $^2$Institute for Theoretical and Mathematical Physics,}
\centerline{ \it Lomonosov Moscow State University, 119991 Moscow, Russia}
\centerline{\it $^3$NRC “Kurchatov Institute”, 123182, Moscow, Russia}
\centerline{\it $^4$Institute for Theoretical and Experimental Physics, 117218, Moscow, Russia}
\centerline{\it $^5$Institute for Information Transmission Problems, 127051, Moscow, Russia}

\bigskip

\centerline{ABSTRACT}

\bigskip
{\footnotesize
We revisit a long-standing problem of correlation functions with the Harish-Chandra-Itzykson-Zuber measure.
The immediate goal is to reproduce the long-known answer for {\it pair} correlator in the form of character expansion --a sum over Schur polynomials with carefully adjusted coefficients. The infinite sum can be folded up into exponentials with the help of the Cauchy identity.
What matters are the coefficients, which we express in terms of projectors, made from Cherednik operators. Development of character expansion methods for unitary integrals, which are much more difficult than the Hermitian ones, gives a hope for further generalizations -- up to arbitrary correlators with non-trivial measures.
}

\bigskip

\bigskip

\section{Introduction and discussion}
Calculation of the Itzykson-Zuber (IZ) correlators is a long-standing problem that has been approached in different ways \cite{9209074, Sha, Eynard2004, EyFe}. A particularly interesting approach is through character expansion \cite{MS1}. This expansion was already known for the IZ integral itself \cite{HC,Itzykson3} through Schur polynomials \cite{Kazakov2000,Balantekin2000,0906.3518,MO}, and an interesting problem was to extend it for the correlator, where it's not symmetric in eigenvalues. Recently, several attempts have been made to address this problem in \cite{MS1} using derivatives of Schur polynomials. In this paper, we propose a character expansion for the IZ pair correlator at arbitrary matrix size $N$. The answer is simple, elegant, and suggestive of further generalizations. It agrees with the original results but takes a very different form. The route by which we found this
expression was rather lengthy and will be described elsewhere \cite{MS2}. Here we simply present the answer and demonstrate the matching.\\
\newline
We begin with the character expansion of the IZ integral. For two Hermitian $N\times N$ matrices $X$ and $Y$, the IZ integral is
\begin{equation}
Z_{\text{IZ}}(X,Y)
= \int_{U(N)}dU\,
e^{\operatorname{tr}(XUYU^\dagger)}, \qquad \int_{U(N)}dU=1
\label{IZRR}
\end{equation}
It depends only on the eigenvalues $x_i,y_j$ and
admits the exact evaluation
\cite{Itzykson3,HC}
\begin{equation}
Z_N(X,Y)
= \left(\prod_{r=1}^{N-1}r!\right)
\frac{\det_{1\leq i,j\leq N}(e^{x_i y_j})} {\Delta(X)\Delta(Y)}, \qquad
\Delta(X)=\prod_{i<j}(x_i-x_j)
\end{equation}
A character expansion expresses the integrand as a sum of characters, which are traces in irreducible representations. For the polynomial representations of $GL(N)$, labeled by partitions $\lambda$ with $\ell(\lambda)\leq N$, these characters are Schur polynomials. To obtain the character expansion of the IZ integral, we use
the well-known Cauchy identity \cite{0906.3518,Cauchy}
\begin{equation}
\sum_\lambda
S_\lambda\{p\}S_\lambda\{\bar p\} = \exp\left(
\sum_{r\geq1}\frac{p_r\bar p_r}{r}
\right)
\end{equation}
Setting
$p_r=\tr[(XUYU^\dagger)^r]$
and $\bar p_r=\delta_{r,1}$ gives
\begin{equation}
e^{\tr(XUYU^\dagger)}
= \sum_{\ell(\lambda)\leq N} S_\lambda\{\delta_{r,1}\} S_\lambda(XUYU^\dagger)
\end{equation}
The character interpretation now allows us to perform the unitary average. Haar orthogonality gives
\begin{equation}
\begin{aligned}
\int dU\,S_\lambda(XUYU^\dagger)
&=
\int dU\,
\operatorname{Tr}_{V_\lambda}
\left(
\rho_\lambda(X)\rho_\lambda(U)
\rho_\lambda(Y)\rho_\lambda(U)^\dagger
\right) = \frac{S_\lambda(X)S_\lambda(Y)}{S_\lambda[N]}
\end{aligned}
\end{equation}
where
$S_\lambda[N]=\dim V_\lambda$. Substitution this into \eqref{IZRR} therefore gives the following character expansion of the IZ integral, where the numerator of the coefficient comes from the Cauchy expansion, while the denominator comes from Haar orthogonality:
\begin{equation}
    Z_{IZ}[X,Y] = \int_{N \times N} e^{\tr (XUYU^\dagger )}[dU] =  \sum_{\lambda} \frac{S_\lambda\{ \delta_{k,1} \} S_\lambda(X) S_\lambda(Y)}{S_\lambda[N]}
    \label{IZR}
\end{equation}
Character expansion provides an algebraic approach to unitary matrix integrals  \cite{0906.3518,Balantekin2000}. It organizes each degree into a finite sum over Young diagrams, makes the matrix-size dependence explicit, and provides a representation-theoretic starting point for studying more general insertions. It is therefore natural to try this approach for the IZ correlators. \\
\newline
The central question that we answer in this paper is to shortly explain how the pair insertion changes this character expansion \eqref{IZR} and show explicitly, for $N=2$, how the resulting infinite series sums to the known exponential expressions\cite{9209074}. Perhaps, we organize the presentation of this paper in the following way:

\begin{itemize}
\item
We address the difficulties of the character expansion of the IZ correlator due to a peculiar feature.
\item
We present our main result, the Schur formula for the pair correlator at arbitrary $N$ and explain its notation and the role of Cherednik projectors \cite{Forrester1}.
\item
We illustrate the formula for $N=2$ and sum the series to recover the known exponential expressions from\cite{9209074}. We also briefly discuss the checks at $N=3$.
\end{itemize}

\section{From the IZ character expansion to the pair correlator}

The Cauchy identity gives a simple way to expand the exponential in the IZ integral. We may apply the same identity in the presence of a pair insertion:
\begin{equation}
\left\langle U_{ij}U^\dagger_{kl}\right\rangle_{\text{IZ}}
=
\sum_{\ell(\lambda)\leq N}
S_\lambda\{\delta_{r,1}\}
\int dU\,U_{ij}U^\dagger_{kl}
S_\lambda(XUYU^\dagger)
\label{pair-cauchy}
\end{equation}
The difficulty is therefore not with the Cauchy identity itself, but with the unitary average that remains. With the additional factor $U_{ij}U^\dagger_{kl}$, this identity
no longer evaluates the required average. Thus the Cauchy expansion alone does not give an explicit Schur formula for the pair correlator.\\
\newline
We may instead try to obtain the pair correlator by differentiating the Cauchy identity. For diagonal $X$ and $Y$, we set $p_r=\operatorname{tr}X^r$ and
$\bar p_r=\operatorname{tr}Y^r$. Taking one derivative with respect to $x_i$ and another with respect to $y_j$, we obtain
\begin{equation}
\begin{aligned}
\sum_\lambda
\frac{\partial S_\lambda(X)}{\partial x_i}
\frac{\partial S_\lambda(Y)}{\partial y_j}
= \frac{\partial^2}{\partial x_i\,\partial y_j}
\exp\left(\sum_{r\geq1}
\frac{\operatorname{tr}X^r\operatorname{tr}Y^r}{r}
\right) = \frac{\partial^2}{\partial x_i\,\partial y_j} \prod_{a,b=1}^{N}\frac{1}{1-x_a y_b}
\end{aligned}
\end{equation}
The differentiated Cauchy identity suggests using products of Schur derivatives to express the pair correlator. For the IZ integral, these products carry coefficients supplied by unitary integration. By differentiating the integral formula and the character expansion of the IZ partition function, we get
\begin{equation}
\frac{\partial^2 Z_{\text{IZ}}}{\partial Y_{jk}\,\partial X_{li}}
= \bigl\langle U_{ij}U^\dagger_{kl}\bigr\rangle_{\text{IZ}} + \sum_{a,b,c,d=1}^{N}Y_{ab}X_{cd} \bigl\langle U_{ia}U^\dagger_{bl}U^\dagger_{kc}U_{dj} \bigr\rangle_{\text{IZ}} = \sum_\lambda \frac{S_\lambda\{ \delta_{k,1} \}}{S_\lambda[N]}\frac{\partial S_\lambda(X)}{\partial X_{ij}} \frac{\partial S_\lambda(Y)}{\partial Y_{kl}}
\label{IZR3}
\end{equation}
 The IZ model, however, has this peculiar feature. The products of Schur derivatives in \eqref{IZR3} describe the pair correlator together with a weighted sum of four-point correlators. They do not isolate the pair contribution. This motivates us to look for a different way to combine the Schur derivatives. In the next section, we present our formula, which decomposes these derivatives into components associated with the removable corners of each Young diagram.

\section{Schur formula for the IZ pair correlator}
We propose the following Schur formula for the IZ pair correlator at arbitrary matrix size $N$:
\begin{equation}
\boxed{
\begin{aligned}
\left\langle U_{ij}U^\dagger_{kl}\right\rangle_{\text{IZ}}
=\delta_{il}\delta_{jk} \sum_{\substack{\lambda\ne\varnothing\\\ell(\lambda)\le N}}
\sum_{\mu\nearrow\lambda}
\frac{\bigl(S_\lambda\{\delta_{r,1}\}\bigr)^2} {S_\lambda[N]\,S_\mu\{\delta_{r,1}\}} \mathcal{P}_{\lambda,\mu}^{(i)}(X) \mathcal{P}_{\lambda,\mu}^{(i)}(Y)
\end{aligned}}
\label{master201}
\end{equation}
where the projectors
\begin{equation}
\boxed{
\mathcal{P}_{\lambda,\mu}^{(i)}(X)
= \left[ \left( \prod_{\substack{\nu\nearrow\lambda\\\nu\ne\mu}}
\frac{\mathcal J_X-c_{\lambda/\nu}I} {c_{\lambda/\mu}-c_{\lambda/\nu}} \right)
\frac{\partial S_\lambda(X)}{\partial X_{ii}}
\right] , \qquad \mathcal{P}_{\lambda,\mu}^{(i)}(Y) =  \left[ \left( \prod_{\substack{\nu\nearrow\lambda\\\nu\ne\mu}}
\frac{\mathcal J_Y-c_{\lambda/\nu}I} {c_{\lambda/\mu}-c_{\lambda/\nu}}
\right) \frac{\partial S_\lambda(Y)}{\partial Y_{jj}}
\right]}
\end{equation}
where the outer sum in \eqref{master201} runs over nonempty Young diagrams $\lambda$ with at
most $N$ rows. We write $\mu\nearrow\lambda$ when $\mu$ is obtained by removing one corner box from $\lambda$. If that box is in row $a$ and column $b$, its content is $c_{\lambda/\mu}=b-a$. The operators $\mathcal J_X$ and $\mathcal J_Y$ are the Cherednik operators \cite{Che, Forrester1} associated with $X_{ii}$ and $Y_{jj}$, respectively and $I$ is an identity operator. For a fixed $\mu$, the product runs over all other corner removals $\nu$ and selects the component of the Schur derivative associated with $\mu$. The two brackets use the same removal $\mu$. If $\lambda$ has only one removable corner, the product is empty and the full Schur derivative remains. Here, the general form of this Cherednik operator that acts on the Schur derivatives is
\begin{equation}
\begin{aligned}
\mathcal J_X =
X_{ii}\frac{\partial}{\partial X_{ii}}
+\sum_{\substack{a=1\\a\ne i}}^N
\frac{X_{aa}}{X_{ii}-X_{aa}}
\left(I-P_{ia}^{(X)}\right),
\qquad
\mathcal J_Y =
Y_{jj}\frac{\partial}{\partial Y_{jj}}
+\sum_{\substack{b=1\\b\ne j}}^N
\frac{Y_{bb}}{Y_{jj}-Y_{bb}}
\left(I-P_{jb}^{(Y)}\right)
\end{aligned}
\label{ced80}
\end{equation}
Here $P_{ia}^{(X)}$ exchanges $X_{ii}$ and $X_{aa}$, while $P_{jb}^{(Y)}$ exchanges $Y_{jj}$ and $Y_{bb}$. The Cherednik operator \cite{Forrester1} distinguishes the components of a Schur derivative by their eigenvalues. Each component is associated with a removable corner of $\lambda$, and its eigenvalue is the content of that corner. The projector factors in our formula
use these eigenvalues to select one component at a time. For example, take $N=2$, $i=1$, and $\lambda=(2,1)$. The Cherednik operator become
\begin{equation}
\mathcal J_X
= x_1\frac{\partial}{\partial x_1}
  +\frac{x_2}{x_1-x_2}\left(I-\mathsf P_{12}\right),
\label{cherednik}
\end{equation}
where $\mathsf P_{12}$ exchanges $x_1$ and $x_2$. We have
\begin{equation}
F(X)=\frac{\partial S_{\ydiagram{2,1}}(X)}{\partial X_{11}}
=2x_1x_2+x_2^2,
\qquad
\mathcal J_XF(X)=x_1x_2-x_2^2
\end{equation}
The two removable corners have contents
$c_{(2,1)/(1,1)}=1$ and $c_{(2,1)/(2)}=-1$.
The corresponding projected components are
\begin{equation}
\begin{aligned}
&\mathcal P_{(2,1),(1,1)}^{(1)}(X)
=\frac{\mathcal J_X+I}{2}F(X)
=\frac32x_1x_2
\\&
\mathcal P_{(2,1),(2)}^{(1)}(X)
=\frac{I-\mathcal J_X}{2}F(X)
=\frac12x_1x_2+x_2^2
\end{aligned}
\end{equation}
Their sum reproduces the full Schur derivative:
\begin{equation}
F(X)=\mathcal P_{(2,1),(1,1)}^{(1)}(X) + \mathcal P_{(2,1),(2)}^{(1)}(X)
\end{equation}
and the Cherednik operator satisfies
\begin{equation}
    \mathcal J_X \mathcal P_{(2,1),(1,1)}^{(1)}(X) =  \mathcal P_{(2,1),(1,1)}^{(1)}(X), \qquad \mathcal J_X \mathcal P_{(2,1),(2)}^{(1)}(X) =  -\mathcal P_{(2,1),(2)}^{(1)}(X)
\end{equation}
Thus $(\mathcal J_X+I)/2$ preserves $\mathcal P_{(2,1),(1,1)}^{(1)}(X)$ and annihilates $\mathcal P_{(2,1),(2)}^{(1)}(X)$, while $(I-\mathcal J_X)/2$ preserves $\mathcal P_{(2,1),(2)}^{(1)}(X)$ and annihilates $\mathcal P_{(2,1),(1,1)}^{(1)}(X)$. Applying the same projectors to the $Y$-derivative, the contribution of $\lambda=(2,1)$ to $\langle U_{11}U^\dagger_{11}\rangle_{\text{IZ}}$ is
\begin{equation}
\begin{aligned}
\left.\left\langle U_{11}U^\dagger_{11}\right\rangle\right|^{N=2}_3 = \frac{1}{72} \frac{\partial S_{\ydiagram{3}}(x)}{\partial x_1}
\frac{\partial S_{\ydiagram{3}}(y)}{\partial y_1}+\frac{1}{9}\left( P_{(2,1),(1,1)}^{(1)}(X)  P_{(2,1),(1,1)}^{(1)}(Y)  + P_{(2,1),(2)}^{(1)}(X) P_{(2,1),(2)}^{(1)}(Y) \right)
\end{aligned}
\label{qq5}
\end{equation}
The Cherednik operator separates each Schur derivative into components $\mathcal P_{\lambda,\mu}^{(i)}(X)$ associated with the removable corners of $\lambda$. Our formula pairs components corresponding to the same remaining diagram $\mu$, through products $\mathcal P_{\lambda,\mu}^{(i)}(X)\mathcal P_{\lambda,\mu}^{(j)}(Y)$. Multiplying the full Schur derivatives would also produce mixed products
$\mathcal P_{\lambda,\mu}^{(i)}(X)\mathcal P_{\lambda,\nu}^{(j)}(Y)$ with $\mu\ne\nu$, which are absent from our formula. If $\lambda$ has only one removable corner, the projector is the identity
and the component is the full Schur derivative.\\
\newline
Now, expanding the sums in \eqref{master201} into several Young diagrams gives
\begin{equation}
\begin{aligned}
&\left\langle U_{ij}U^\dagger_{kl}\right\rangle_{\text{IZ}}^{N=2} =\delta_{il}\delta_{jk}\Biggl\{
 \frac12
 \frac{\partial S_{\ydiagram{1}}(X)}{\partial X_{ii}}
 \frac{\partial S_{\ydiagram{1}}(Y)}{\partial Y_{jj}} +\frac1{12}
 \frac{\partial S_{\ydiagram{2}}(X)}{\partial X_{ii}}
 \frac{\partial S_{\ydiagram{2}}(Y)}{\partial Y_{jj}}
 +\frac14
 \frac{\partial S_{\ydiagram{1,1}}(X)}{\partial X_{ii}}
 \frac{\partial S_{\ydiagram{1,1}}(Y)}{\partial Y_{jj}} + \\
&\qquad +\frac1{72}
 \frac{\partial S_{\ydiagram{3}}(X)}{\partial X_{ii}}
 \frac{\partial S_{\ydiagram{3}}(Y)}{\partial Y_{jj}}
 +\left(
 \frac19\frac{\mathcal J_X+I}{2}\frac{\mathcal J_Y+I}{2}
 +\frac19\frac{I-\mathcal J_X}{2}\frac{I-\mathcal J_Y}{2}
 \right)
 \left(
 \frac{\partial S_{\ydiagram{2,1}}(X)}{\partial X_{ii}}
 \frac{\partial S_{\ydiagram{2,1}}(Y)}{\partial Y_{jj}}
 \right)
\\
&\qquad
 +\frac1{480}
 \frac{\partial S_{\ydiagram{4}}(X)}{\partial X_{ii}}
 \frac{\partial S_{\ydiagram{4}}(Y)}{\partial Y_{jj}}
 +\frac1{48}
 \frac{\partial S_{\ydiagram{2,2}}(X)}{\partial X_{ii}}
 \frac{\partial S_{\ydiagram{2,2}}(Y)}{\partial Y_{jj}}
\\
&\qquad
 +\left( \frac1{64}\frac{\mathcal J_X+I}{3}\frac{\mathcal J_Y+I}{3}
 +\frac1{32}\frac{2I-\mathcal J_X}{3}\frac{2I-\mathcal J_Y}{3}
 \right) \left( \frac{\partial S_{\ydiagram{3,1}}(X)}{\partial X_{ii}}
 \frac{\partial S_{\ydiagram{3,1}}(Y)}{\partial Y_{jj}}
 \right)
\\
&\qquad
 +\frac1{3600} \frac{\partial S_{\ydiagram{5}}(X)}{\partial X_{ii}} \frac{\partial S_{\ydiagram{5}}(Y)}{\partial Y_{jj}}
\\
&\qquad
 +\left( \frac1{450}\frac{\mathcal J_X+I}{4}\frac{\mathcal J_Y+I}{4}
 +\frac1{150}\frac{3I-\mathcal J_X}{4}\frac{3I-\mathcal J_Y}{4} \right)
 \left( \frac{\partial S_{\ydiagram{4,1}}(X)}{\partial X_{ii}} \frac{\partial S_{\ydiagram{4,1}}(Y)}{\partial Y_{jj}} \right)
\\
&\qquad
 +\left( \frac1{96}\frac{\mathcal J_X}{2}\frac{\mathcal J_Y}{2}
 +\frac1{144}\frac{2I-\mathcal J_X}{2}\frac{2I-\mathcal J_Y}{2}
 \right) \left( \frac{\partial S_{\ydiagram{3,2}}(X)}{\partial X_{ii}}\frac{\partial S_{\ydiagram{3,2}}(Y)}{\partial Y_{jj}}\right)
 +\cdots \Biggr\}
\end{aligned}
\label{oo1}
\end{equation}
We now evaluate the Young diagram sum for $\langle U_{11}U^\dagger_{11}\rangle_{\text{IZ}}$ at $N=2$, using the explicit Schur polynomials and the projectors above. We collect the contributions from all diagrams with $m+1$
boxes and denote their sum by $G_{2m}$. Each such contribution has degree $m$ in the $x$ eigenvalues and degree $m$ in the $y$ eigenvalues, giving total degree $2m$. Thus,
\begin{equation}
\left\langle U_{11}U^\dagger_{11}\right\rangle_{\text{IZ}}
=\sum_{m=0}^{\infty}G_{2m}
\end{equation}
The first five terms, obtained from diagrams with up to five boxes, are
\begin{equation}
    \begin{aligned}
        G_0 =& \frac{1}{2}\\
        G_2 =& \frac{1}{6} \left(x_1 \left(2 y_1+y_2\right)+x_2 \left(y_1+2 y_2\right)\right)\\
        G_4 =& \frac{1}{24} \left(x_1^2 \left(3 y_1^2+2 y_2 y_1+y_2^2\right)+2 x_2 x_1 \left(y_1^2+4 y_2 y_1+y_2^2\right)+x_2^2
   \left(y_1^2+2 y_2 y_1+3 y_2^2\right)\right)\\
   G_6 =& \frac{1}{120} \Bigl(x_1^3 \left(4 y_1^3+3 y_2 y_1^2+2 y_2^2 y_1+y_2^3\right) + x_2 x_1^2 \left(3 y_1^3+16 y_2 y_1^2+9 y_2^2 y_1+2 y_2^3\right) \\ & + x_2^2 x_1 \left(2 y_1^3+9 y_2 y_1^2+16 y_2^2 y_1+3 y_2^3\right) + x_2^3 \left(y_1^3+2 y_2 y_1^2+3 y_2^2 y_1+4 y_2^3\right) \Bigr)\\
   G_8 =& \frac{1}{720} \Bigl(x_1^4 \left(5 y_1^4+4 y_2 y_1^3+3 y_2^2 y_1^2+2 y_2^3 y_1+y_2^4\right) + 2 x_2 x_1^3 \left(2 y_1^4+13 y_2 y_1^3+9 y_2^2 y_1^2+5 y_2^3 y_1+y_2^4\right) \\ & + 3 x_2^2 x_1^2 \left(y_1^4+6 y_2 y_1^3+16 y_2^2 y_1^2+6 y_2^3 y_1+y_2^4\right) + 2 x_2^3 x_1 \left(y_1^4+5 y_2 y_1^3+9 y_2^2 y_1^2+13 y_2^3 y_1+2 y_2^4\right) \\ & + x_2^4 \left(y_1^4+2 y_2 y_1^3+3 y_2^2 y_1^2+4 y_2^3 y_1+5 y_2^4\right) \Bigr)
\end{aligned}
\label{sumz}
\end{equation}

\section{Consistency with the known formulas}
We now show how the Schur series obtained above sums to the known exponential expressions for the pair correlator. For $N=2$, with $X_{mn}=x_m-x_n$ and $Y_{rs}=y_r-y_s$, the IZ pair correlator in equation (8) of \cite{9209074} read
\begin{equation}
\braket{U_{11}U_{11}^\dagger}
=\braket{U_{22}U_{22}^\dagger}
=\frac{e^{x_1y_1+x_2y_2}}{X_{12}Y_{12}}
 \left(1-\frac{1}{X_{12}Y_{12}}\right)
+\frac{e^{x_1y_2+x_2y_1}}{(X_{12}Y_{12})^2}
\label{pair11}
\end{equation}
\begin{equation}
\braket{U_{12}U_{21}^\dagger}
=\braket{U_{21}U_{12}^\dagger}
=\frac{e^{x_1y_1+x_2y_2}}{(X_{12}Y_{12})^2}
-\frac{e^{x_1y_2+x_2y_1}}{X_{12}Y_{12}}
 \left(1+\frac{1}{X_{12}Y_{12}}\right)
 \label{pair22}
\end{equation}
To compare these expressions with our Schur expansion \eqref{oo1}, we return to the terms $G_0,\ldots,G_8$ calculated above \eqref{sumz}.
Their structure becomes simpler when we group the eigenvalues into the combinations $x_1y_1+x_2y_2$ and $x_1y_2+x_2y_1$. Rewriting the first few terms in this form reveals a pattern that suggests how to sum the series. For this calculation, we write
\begin{equation}
a=x_1y_1+x_2y_2,
\qquad
b=x_1y_2+x_2y_1,
\qquad
a-b=X_{12}Y_{12}
\end{equation}
The terms calculated above \eqref{sumz} can then be written as
\begin{equation}
\begin{aligned}
& G_0=\frac{1}{2!}, \qquad G_2=\frac{2a+b}{3!}, \qquad G_4=\frac{3a^2+2ab+b^2}{4!}, \\[0.5cm]& G_6=\frac{4a^3+3a^2b+2ab^2+b^3}{5!}, G_8=\frac{5a^4+4a^3b+3a^2b^2+2ab^3+b^4}{6!}
\end{aligned}
\end{equation}
At each grading, the coefficients decrease by one as the power of $b$ increases. These terms suggest the general expression
\begin{equation}
G_{2m} = \frac{1}{(m+2)!} \sum_{k=0}^{m}(m+1-k)a^{m-k}b^k
\end{equation}
We can now sum this series explicitly. The coefficient $m+1-k$ appears when we differentiate $a^{m+1-k}$ with respect to $a$. Therefore, keeping $b$ fixed,
\begin{equation}
G_{2m}
= \frac{1}{(m+2)!}
\frac{\partial}{\partial a}
\left(a^{m+1}+a^m b+\cdots+ab^m+b^{m+1}
\right)
\end{equation}
We have included the last term $b^{m+1}$ because its $a$-derivative vanishes. The expression in parentheses is a finite geometric sum:
\begin{equation}
a^{m+1}+a^m b+\cdots+ab^m+b^{m+1} = \frac{a^{m+2}-b^{m+2}}{a-b}
\end{equation}
Summing over all gradings consequently gives
\begin{equation}
\sum_{m=0}^{\infty}G_{2m} = \frac{\partial}{\partial a} \left[ \frac{1}{a-b} \sum_{m=0}^{\infty} \frac{a^{m+2}-b^{m+2}}{(m+2)!}
\right]
\end{equation}
The factorial series starts at the quadratic term. Using the Taylor expansion of the exponential and setting $n=m+2$, we obtain:
\begin{equation}
\begin{aligned}
\sum_{m=0}^{\infty}
\frac{a^{m+2}-b^{m+2}}{(m+2)!}
=(e^a-1-a)-(e^b-1-b) =e^a-e^b-(a-b)
\end{aligned}
\end{equation}
It follows that
\begin{equation}
\begin{aligned}
\sum_{m=0}^{\infty}G_{2m} = \frac{\partial}{\partial a} \left( \frac{e^a-e^b}{a-b}-1 \right) =\frac{e^a}{a-b} -\frac{e^a-e^b}{(a-b)^2}
\end{aligned}
\end{equation}
Restoring the eigenvalues, we obtain
\begin{equation}
\left\langle U_{11}U^\dagger_{11}\right\rangle_{\text{IZ}}
= \frac{e^{x_1y_1+x_2y_2}}{X_{12}Y_{12}}
\left(1-\frac{1}{X_{12}Y_{12}}\right)
+ \frac{e^{x_1y_2+x_2y_1}}{(X_{12}Y_{12})^2}
\end{equation}
This is the correlator in equation (8) of
\cite{9209074}. The two exponentials arise from summing the factorial series in $a$ and $b$, while the factors involving $X_{12}Y_{12}$ arise from the finite geometric sum and its derivative. Exchanging $y_1$ and $y_2$ gives the other set of correlator. This exchange interchanges $a$ and $b$
and changes the sign of $Y_{12}$, yielding
\begin{equation}
\left\langle U_{12}U^\dagger_{21}\right\rangle_{\text{IZ}}
= \frac{e^{x_1y_1+x_2y_2}}{(X_{12}Y_{12})^2}
- \frac{e^{x_1y_2+x_2y_1}}{X_{12}Y_{12}}
\left(1+\frac{1}{X_{12}Y_{12}}\right)
\end{equation}
We next consider $N=3$. Our proposed formula \eqref{master201} has the same structure, with the coefficients now determined by $S_\lambda[3]$ and the sum extending over diagrams with at most three rows. The corner contents remain unchanged, so the projectors for diagrams already present at $N=2$ can be used again. The additional three-row diagrams contribute according to the same prescription. Keeping all diagrams with up to four boxes, we obtain
\begin{equation}
\begin{aligned}
&\left\langle U_{ij}U^\dagger_{kl}\right\rangle_{\text{IZ}}^{N=3} =\delta_{il}\delta_{jk}\Biggl\{
 \frac13 \frac{\partial S_{\ydiagram{1}}(X)}{\partial X_{ii}}\frac{\partial S_{\ydiagram{1}}(Y)}{\partial Y_{jj}} +\frac1{24} \frac{\partial S_{\ydiagram{2}}(X)}{\partial X_{ii}}\frac{\partial S_{\ydiagram{2}}(Y)}{\partial Y_{jj}} +\frac1{12} \frac{\partial S_{\ydiagram{1,1}}(X)}{\partial X_{ii}} \frac{\partial S_{\ydiagram{1,1}}(Y)}{\partial Y_{jj}}
 +\\
&\qquad + \frac1{180} \frac{\partial S_{\ydiagram{3}}(X)}{\partial X_{ii}} \frac{\partial S_{\ydiagram{3}}(Y)}{\partial Y_{jj}} +\frac1{18} \frac{\partial S_{\ydiagram{1,1,1}}(X)}{\partial X_{ii}}\frac{\partial S_{\ydiagram{1,1,1}}(Y)}{\partial Y_{jj}}
\\
&\qquad
 +\left( \frac1{36}\frac{\mathcal J_X+I}{2}\frac{\mathcal J_Y+I}{2} +\frac1{36}\frac{I-\mathcal J_X}{2}\frac{I-\mathcal J_Y}{2} \right)
 \left(\frac{\partial S_{\ydiagram{2,1}}(X)}{\partial X_{ii}} \frac{\partial S_{\ydiagram{2,1}}(Y)}{\partial Y_{jj}} \right)
\\
&\qquad
 +\frac{1}{1440} \frac{\partial S_{\ydiagram{4}}(X)}{\partial X_{ii}} \frac{\partial S_{\ydiagram{4}}(Y)}{\partial Y_{jj}} +\frac{1}{288} \frac{\partial S_{\ydiagram{2,2}}(X)}{\partial X_{ii}} \frac{\partial S_{\ydiagram{2,2}}(Y)}{\partial Y_{jj}}
\\
&\qquad
 +\left(
   \frac{1}{320}
   \frac{\mathcal J_X+I}{3}\frac{\mathcal J_Y+I}{3} +\frac{1}{160}
   \frac{2I-\mathcal J_X}{3}\frac{2I-\mathcal J_Y}{3} \right)
  \left( \frac{\partial S_{\ydiagram{3,1}}(X)}{\partial X_{ii}} \frac{\partial S_{\ydiagram{3,1}}(Y)}{\partial Y_{jj}}
  \right)
\\
&\qquad
 +\left( \frac{1}{32}
   \frac{\mathcal J_X+2I}{3}\frac{\mathcal J_Y+2I}{3} +\frac{1}{64}
   \frac{I-\mathcal J_X}{3}\frac{I-\mathcal J_Y}{3}  \right) \left( \frac{\partial S_{\ydiagram{2,1,1}}(X)}{\partial X_{ii}} \frac{\partial S_{\ydiagram{2,1,1}}(Y)}{\partial Y_{jj}} \right) +\cdots
 \Biggr\}
\end{aligned}
\label{oo2}
\end{equation}
We compare this expansion with the known $N=3$ correlators in equation (13)-(15) of \cite{9209074}. We evaluate these projected Schur derivatives \eqref{oo2} and compare the resulting polynomials with the exponential expansion in\cite{9209074}. The contributions from all diagrams with up to four boxes agree through total degree six. This includes the additional three-row diagrams that first appear at $N=3$. The remaining pair correlators follow by relabeling the eigenvalues. These checks show that the same prescription for the coefficients and corner projectors continues to reproduce
the known correlators at $N=3$. Together with the $N=2$ resummation, they provide further support for \eqref{master201} at arbitrary $N$.

\section{Concluding remarks}

We found and discussed a character expansion formula (\ref{master201}) for pair correlators in the Harish-Chandra-Itzykson-Zuber unitary-matrix integral \cite{HC,Itzykson3}. It involves somewhat complicated projectors, made from Cherednik operators \cite{Che, Forrester1}, and can look far more overloaded than the original explicit expressions in \cite{9209074} and \cite{Sha,Eynard2004,EyFe}. However, they are conceptually important, because character expansion seems to be a general method \cite{0906.3518}, applicable in a systematic way to arbitrary quantum field theory models, and allowing to sum up perturbation series with the help of Cauchy-like identities. Understanding of these basic sum rules and their applications can be important for development of generic non-perturbative calculus, and the work on unitary integrals is an important step on this way. The knowledge of simple answers, to be reproduced by this method, is very helpful. Their lack for higher IZ correlators is an important difficulty, which we now hope to overcome, after the relevant structures are revealed and analyzed. \\
\newline
In \cite{MS2} we provide a description of various attempts, which were not all successful, but finally led us to (\ref{master201}) -- some of abandoned attempts can be actually developed further, extending the existing insights. A possible simplification of our formulas in \cite{Drachov} 
in terms of non-symmetric polynomials {\it a la} \cite{2512.24811} 
can also help on this way --
towards character expansion for generic IZ correlators and, further, to non-pertutbative IZ-based unitary partition function, which can complement and extend the knowledge, already existing about the simplest unitary Brezin-Gross-Witten (BGW) model \cite{BGW1,BGW2,MMS}.

\newpage
\section*{Acknowledgements}
We are grateful to Yaroslav Drachov, Aleksandr Popolitov, Victor Mishnyakov, Anastasia Oreshina, Pavel Suprun and Maxim Reva for many illuminating discussions about this study. The work was supported by the state assignment of the Institute for Information Transmission Problems of RAS. HS was also supported by the scholarship from the Theoretical Physics and Mathematics Advancement Foundation «BASIS». \\
We acknowledge ChatGPT Sol and Astra for minor computations and cross-checking some of our expressions. All results were verified by the authors, who take full responsibility for the content.


\begin{thebibliography}{12}
\bibitem{9209074}
A. Morozov,
“Pair correlator in the Itzykson--Zuber integral'',
\href{https://doi.org/10.1142/S0217732392002913}
{\emph{Mod. Phys. Lett. A} \textbf{7}, no.~37 (1992) 3503--3507},
\href{https://arxiv.org/abs/hep-th/9209074}{arXiv:hep-th/9209074}.

\bibitem{Sha}
S.L. Shatashvili,
“Correlation functions in the Itzykson--Zuber model'',
\href{https://doi.org/10.1007/BF02097004}
{\emph{Commun. Math. Phys.} \textbf{154} (1993) 421--432},
\href{https://arxiv.org/abs/hep-th/9209083}{arXiv:hep-th/9209083}.

\bibitem{Eynard2004}
B. Eynard,
“A short note about Morozov's formula'',
\href{https://arxiv.org/abs/math-ph/0406063}{arXiv:math-ph/0406063}.

\bibitem{EyFe}
B. Eynard and A. Prats Ferrer,
“2-matrix versus complex matrix model, integrals over the unitary group as triangular integrals'',
\href{https://doi.org/10.1007/s00220-006-1541-8}
{\emph{Commun. Math. Phys.} \textbf{264} (2006) 115--144},
\href{https://arxiv.org/abs/hep-th/0502041}{arXiv:hep-th/0502041}.

\bibitem{MS1}
A.Morozov, H.Sifat, “Itzykson-Zuber correlators from character expansion”, \href{https://arxiv.org/abs/2504.05292}{\emph{Eur. Phys. J. C} 85 (2025) \textbf{648}, arXiv:2504.05292}.

\bibitem{HC}
Harish-Chandra,
“Differential operators on a semisimple Lie algebra'',
\href{https://doi.org/10.2307/2372387}
{\emph{Amer. J. Math.} \textbf{79} (1957) 87--120}.

\bibitem{Itzykson3}
C. Itzykson and J.-B. Zuber,
“The planar approximation. II'',
\href{https://doi.org/10.1063/1.524438}
{\emph{J. Math. Phys.} \textbf{21} (1980) 411--421}.

\bibitem{Kazakov2000}
V.A. Kazakov,
“Solvable Matrix Models'',
in \emph{Random Matrix Models and Their Applications},
\href{https://doi.org/10.1017/9781009701440.013}
{\emph{Math. Sci. Res. Inst. Publ.} \textbf{40} (2001) 271--284}.

\bibitem{Balantekin2000}
A.B. Balantekin,
\emph{Character expansions, Itzykson--Zuber integrals,
and the QCD partition function},
\href{https://doi.org/10.1103/PhysRevD.62.085017}
{Phys. Rev. D \textbf{62} (2000) 085017}
[\href{https://arxiv.org/abs/hep-th/0007161}{hep-th/0007161}].

\bibitem{0906.3518}
A. Morozov,
\emph{Unitary integrals and related matrix models},
\href{https://doi.org/10.1007/s11232-010-0001-y}
{Theor. Math. Phys. \textbf{162} (2010) 1--33}
[\href{https://arxiv.org/abs/0906.3518}{arXiv:0906.3518}].

\bibitem{MO}
A. Morozov and A. Oreshina,
“On character expansion and Gaussian regularization of Itzykson--Zuber measure'',
\href{https://doi.org/10.1016/j.physletb.2024.139006}
{\emph{Phys. Lett. B} \textbf{857} (2024) 139006},
\href{https://arxiv.org/abs/2405.13579}{arXiv:2405.13579}.

\bibitem{MS2}
A.Morozov, H.Sifat, “Tensor diagrams and Ward identities in unitary matrix models”, \emph{to appear}

\bibitem{Cauchy}
A. Morozov,
“Cauchy formula and the character ring'',
\href{https://doi.org/10.1140/epjc/s10052-019-6598-6}
{\emph{Eur. Phys. J. C} \textbf{79}, no.~1 (2019) 76},
\href{https://arxiv.org/abs/1812.03853}{arXiv:1812.03853}.

\bibitem{Che}
I. Cherednik,
“A unification of Knizhnik--Zamolodchikov and Dunkl operators
via affine Hecke algebras'', \href{https://doi.org/10.1007/BF01243918}
{\emph{Invent. Math.} \textbf{106} (1991) 411--431}.


\bibitem{Forrester1}
P.J. Forrester, D.S. McAnally and Y. Nikoyalevsky,
“On the evaluation formula for Jack polynomials with prescribed symmetry'',
\href{https://doi.org/10.1088/0305-4470/34/41/302}
{\emph{J. Phys. A: Math. Gen.} \textbf{34} (2001) 8407--8424}, \href{https://arxiv.org/abs/math/0107126}{arXiv:math/0107126}.

\bibitem{Drachov} Ya.Drachov et al., {\it to appear}.

\bibitem{2512.24811}
A. Mironov, A. Morozov and A. Popolitov,
“Twisted Cherednik systems and non-symmetric Macdonald polynomials'', \href{https://doi.org/10.48550/arXiv.2512.24811}
{\emph{arXiv:2512.24811} (2025)}.

\bibitem{BGW1}
D.J. Gross and E. Witten,
“Possible third-order phase transition in the large-$N$ lattice gauge theory'', \href{https://doi.org/10.1103/PhysRevD.21.446}
{\emph{Phys. Rev. D} \textbf{21} (1980) 446--453}.

\bibitem{BGW2}
E. Br\'ezin and D.J. Gross,
“The external field problem in the large $N$ limit of QCD'', \href{https://doi.org}
{\emph{Phys. Lett. B} \textbf{97} (1980) 120--124}.


\bibitem{MMS}
A. Mironov, A. Morozov and G.W. Semenoff,
“Unitary Matrix Integrals in the Framework of Generalized Kontsevich Model. I. Brezin-Gross-Witten Model'',
\href{https://doi.org/10.1142/S0217751X96002339}
{\emph{Int. J. Mod. Phys. A} \textbf{11} (1996) 5031--5080}, \href{https://arxiv.org/abs/hep-th/9404005}{arXiv:hep-th/9404005}.


\end{thebibliography}
\end{document}